\documentclass[twocolumn]{jpsj3}
\usepackage{txfonts}
\RequirePackage{graphicx,color}

\title{
Novel Magnetoacoustic Resonance Technique for Exploring Hidden Quadrupoles \\
in a Crystal Field Quartet
}

\author{Mikito Koga$^1$ and Masashige Matsumoto$^2$}

\inst{
$^1$Department of Physics, Faculty of Education, Shizuoka University, Shizuoka
422--8529, Japan \\
$^2$Department of Physics, Faculty of Science, Shizuoka University, Shizuoka
422--8529, Japan \\
}

\abst{
Crystal field quartets with quadrupole degrees of freedom play a crucial role in hidden
ordering systems, as exemplified by CeB$_6$.
We present a novel magnetoacoustic resonance technique that combines acoustically induced
strain fields with a linearly polarized high-frequency microwave field to probe quadrupoles
inherent in the quartet hidden behind magnetic properties.
This method offers the advantage of enabling quantum quadrupole resonance transitions for large
excitation energy gaps within quartet sublevels under a strong magnetic field, which cannot be
achieved by acoustic experiments alone.
Formulating a simultaneous single-phonon--single-photon absorption transition process
using Floquet theory, we demonstrate how the transition probabilities
are affected by changing the propagation direction of a bulk acoustic wave.
The key result is that distinct maxima in transition probabilities, attributed to specific propagation
directions, indicate a characteristic of quadrupole physics and exhibit an abrupt change owing to
an induced ordered moment.
This photon-assisted magnetoacoustic resonance technique will promote a broader range of
applications of acoustic experiments for the study of quadrupole physics.
}

\begin{document}

\maketitle

%%%%%%%%%%%%%%%%%%%%%%%%%%%%%%%%%%%%%%%%%%%%%%%%%%%%%%%%%%%%%%%%%%%%%%%%%%%%%%%%
%Macros
%%%%%%%%%%%%%%%%%%%%%%%%%%%%%%%%%%%%%%%%%%%%%%%%%%%%%%%%%%%%%%%%%%%%%%%%%%%%%%%%
\newcommand{\ds}{\displaystyle}

\renewcommand{\H}{\mathcal{H}}
\newcommand{\br}{{\mbox{\boldmath$r$}}}
\newcommand{\bR}{{\mbox{\boldmath$R$}}}
\newcommand{\bS}{{\mbox{\boldmath$S$}}}
\newcommand{\bk}{{\mbox{\boldmath$k$}}}
\newcommand{\bH}{{\mbox{\boldmath$H$}}}
\newcommand{\bh}{{\mbox{\boldmath$h$}}}
\newcommand{\bJ}{{\mbox{\boldmath$J$}}}
\newcommand{\bI}{{\mbox{\boldmath$I$}}}
\newcommand{\bPsi}{{\mbox{\boldmath$\Psi$}}}
\newcommand{\bpsi}{{\mbox{\boldmath$\psi$}}}
\newcommand{\bPhi}{{\mbox{\boldmath$\Phi$}}}
\newcommand{\bd}{{\mbox{\boldmath$d$}}}
\newcommand{\bG}{{\mbox{\boldmath$G$}}}
\newcommand{\bu}{{\mbox{\boldmath$u$}}}
\newcommand{\be}{{\mbox{\boldmath$e$}}}
\newcommand{\bE}{{\mbox{\boldmath$E$}}}
\newcommand{\bp}{{\mbox{\boldmath$p$}}}
\newcommand{\bB}{{\mbox{\boldmath$B$}}}
\newcommand{\om}{{\omega_n}}
\newcommand{\omm}{{\omega_{n'}}}
\newcommand{\omd}{{\omega^2_n}}
\newcommand{\omt}{{\tilde{\omega}_{n}}}
\newcommand{\ommt}{{\tilde{\omega}_{n'}}}
\newcommand{\brho}{{\mbox{\boldmath$\rho$}}}
\newcommand{\bsigma}{{\mbox{\boldmath$\sigma$}}}
\newcommand{\bSigma}{{\mbox{\boldmath$\Sigma$}}}
\newcommand{\btau}{{\mbox{\boldmath$\tau$}}}
\newcommand{\bfeta}{{\mbox{\boldmath$\eta$}}}
\newcommand{\bskp}{{\mbox{\scriptsize\boldmath $k$}}}
\newcommand{\skp}{{\mbox{\scriptsize $k$}}}
\newcommand{\bsrp}{{\mbox{\scriptsize\boldmath $r$}}}
\newcommand{\bsRp}{{\mbox{\scriptsize\boldmath $R$}}}
\newcommand{\bsk}{\bskp}
\newcommand{\sk}{\skp}
\newcommand{\bsr}{\bsrp}
\newcommand{\bsR}{\bsRp}
\newcommand{\ri}{{\rm i}}
\newcommand{\re}{{\rm e}}
\newcommand{\rd}{{\rm d}}
\newcommand{\rM}{{\rm M}}
\newcommand{\rs}{{\rm s}}
\newcommand{\rt}{{\rm t}}
\newcommand{\Tc}{{$T_{\rm c}$}}
\renewcommand{\Pr}{{PrOs$_4$Sb$_{12}$}}
\newcommand{\La}{{LaOs$_4$Sb$_{12}$}}
\newcommand{\LaPr}{{(La$_{1-x}$Pr${_x}$)Os$_4$Sb$_{12}$}}
\newcommand{\PrLa}{{(Pr$_{1-x}$La${_x}$)Os$_4$Sb$_{12}$}}
\newcommand{\OsRu}{{Pr(Os$_{1-x}$Ru$_x$)$_4$Sb$_{12}$}}
\newcommand{\PrRu}{{PrRu$_4$Sb$_{12}$}}
%%%%%%%%%%%%%%%%%%%%%%%%%%%%%%%%%%%%%%%%%%%%%%%%%%%%%%%%%%%%%%%%%%%%%%%%%%%%%%%%%%%%%%%%%%%%%%%%%%%

%%%%%%%%%%%%%%%%%%%%%%%%%%%%%%%%%%%%%%%%%%%%%%%%%%%%%%%%%%%%%%%%%%%%%%%%%%%%%%%%%%%%%%%%%%%%%%%%%%%
\section{Introduction}
%%%%%%%%%%%%%%%%%%%%%%%%%%%%%%%%%%%%%%%%%%%%%%%%%%%%%%%%%%%%%%%%%%%%%%%%%%%%%%%%%%%%%%%%%%%%%%%%%%%
Crystal field multiplets, characterized by spin $S$ ($ \ge 1$), possess quadrupole degrees of
freedom that can be coupled to strain fields induced by lattice deformations or to
conjugate fields associated with local quadrupolar polarization.
\cite{Nakamura94,Mitsumoto14}
In various magnetic systems such as those involving $f$-electrons, the quadrupole properties,
which are typically hidden behind the magnetic dipole responses, play a significant role in revealing rich phenomena not observed in purely dipolar magnetic systems.
\cite{Santini09,Kuramoto09}
Theoretically, the quadrupoles at atomic sites are represented by rank-2 tensorial forms of local
spin operators linked to charge distribution.
It is crucial to elucidate how the crystal field multiplets exhibit such tensorial quadrupole
characteristics, which remain difficult to access through conventional measurement
techniques used in magnetic materials.
\par

Cerium hexaboride (CeB$_6$) is an exemplary material for investigating the quadrupole physics
associated with localized $f$-electrons.
\cite{Cameron16,Inosov21}
In the Ce $4f^1$ configuration, the sixfold degenerate $f$-electron state with total angular
momentum $J = 5/2$ is stabilized by spin-orbit coupling and is split into $\Gamma_8$ quartet
ground and $\Gamma_7$ doublet excited states within the $O_h$ cubic crystal field environment.
\cite{Lea62}
For the large $\Gamma_8$-$\Gamma_7$ splitting (540~K in energy), one origin of
a hidden order phase (phase II) appearing in CeB$_6$ at 3.4~K can be quadrupole
degrees of freedom inherent in the $\Gamma_8$ quartet.
\cite{Ohkawa85}
The $\Gamma_8$-based scenario for the emergence of an antiferroquadrupolar (AFQ) order was
inferred in earlier experimental studies, including neutron scattering,
\cite{Effantin85}
nuclear magnetic resonance (NMR),
\cite{Takigawa83}
and elastic constant measurements
\cite{Luthi84}.
Notably, the B-site NMR line splitting was successfully explained by group theoretical
studies considering an effect of a $\Gamma_8$ octupole induced by the AFQ
ordered moment in the presence of a magnetic field.
\cite{Shiina97,Sakai97,Shiina98}
The field-induced octupole moment in the AFQ order phase was also
confirmed by the results of resonant X-ray diffraction
\cite{Matsumura12}
and inelastic neutron scattering experiments.
\cite{Portnichenko20}
Above the transition temperature, the robustness of the $\Gamma_8$ symmetry of the Ce $f^1$
ground state was indicated by nonresonant inelastic X-ray scattering.
\cite{Sundermann17}
Although $\Gamma_8$ quadrupoles are crucial for the emergence of the field-induced dipoles
and octupoles, no direct conclusive evidence of the quadrupole dynamics has yet been obtained.
Several issues concerning the local $\Gamma_8$-based scenario have been raised by
electron spin resonance (ESR) studies in high magnetic fields
\cite{Semeno16,Schlottmann18,Semeno21}
and by B-site nuclear quadrupole resonance in very low magnetic fields.
\cite{Mito23}
\par

In a different context, a spin-3/2 quartet can be realized as an electronic ground state of silicon
vacancy (V$_{\rm Si}$) centers in silicon carbide (4H-SiC), and this realization provides an
intriguing platform operating four-level systems as qudits for quantum information
technologies.
\cite{Kraus14,Widmann15,Simin17,Nagy18,Soltamov19,Castelletto20,Son20}
Recent spin-acoustic experiments have revealed that the V$_{\rm Si}$ quartet can be
coupled to strain fields induced by either a surface acoustic wave (SAW) propagating along the
crystal surface or a bulk acoustic wave (BAW).
\cite{Hernandez-Minguez21,Vasselon23,Dietz23}
The acoustically induced strain coupling is attributed to the quadrupole--strain (QS) interaction for
spin-3/2 systems, and its quadrupole nature appears as an anisotropic spin-acoustic resonance
under rotation of the static magnetic field.
\cite{Hernandez-Minguez20,Koga24a}
This observation motivated us to extend our previous study on magnetoacoustic resonance to
uncover microscopic features of quadrupoles inherent in a crystal field quartet.
\cite{Koga24b}
In the present study, in stead of rotating the magnetic field, we rotate the propagation direction of
the acoustic wave so that the Zeeman splitting of the quartet is preserved.
The profiles of quadrupoles can further be captured by measuring the propagation
direction dependence of the acoustically driven quadrupole transition rates between quartet
sublevels.
\par

\begin{figure}
\begin{center}
\includegraphics[width=7cm,clip]{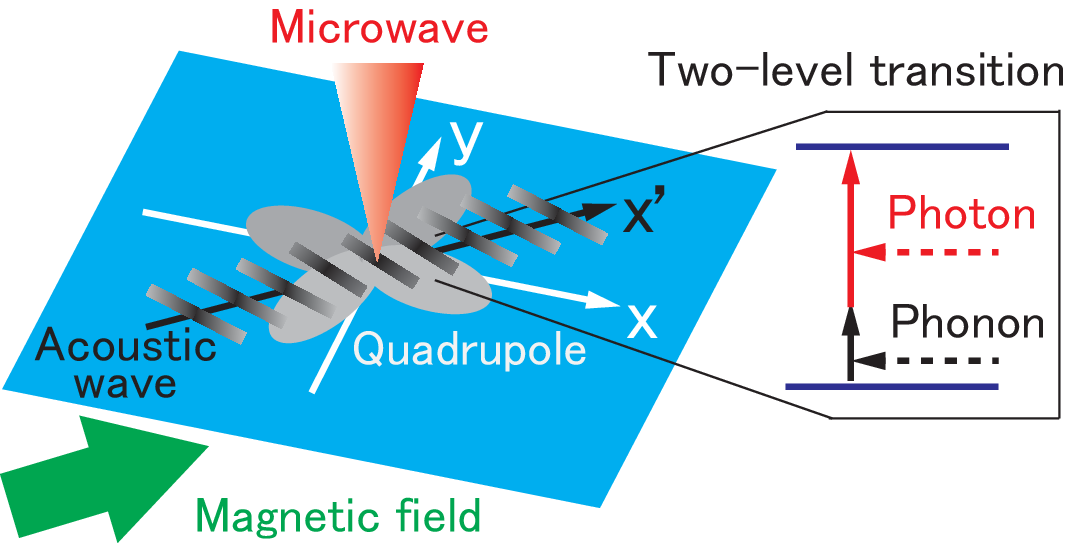}
\end{center}
\caption{
(Color online)
Illustration of photon-assisted magnetoacoustic resonance.
\cite{Yanagisawa24}
The quadrupole resonance transition between two levels of a localized electron state is achieved
through the simultaneous absorption of a single phonon and a single $\pi$-photon
using an acoustic wave and a linearly localized microwave.
The propagation direction ($x'$-axis) of the acoustic wave is rotated in the $xy$ plane under a
static magnetic field.
}
\label{fig:1}
\end{figure}

In conventional methods, there is a difficulty in applying the magnetoacoustic resonance to
the case of large excitation energy gaps in a strong magnetic field ($\sim$1~T), which is inevitably
required for precise measurements.
Usually, an oscillating field of more than 10~GHz order is necessary for a resonance
transition between two levels of a localized electron state.
In contrast, the frequency of an acoustically induced strain field is limited to the gigahertz order.
In a recent study, we proposed photon-assisted magnetoacoustic resonance
(PA-MAR) as an effective method for achieving quadrupole transitions involving such large
excitations.
\cite{Koga24b}
In this method, a relatively low-frequency acoustically induced strain field is combined with a
high-frequency linearly polarized microwave, as shown in Fig.~\ref{fig:1}.
\cite{Yanagisawa24}
Importantly, absorption of the microwave photon ($\pi$-photon) drives only the longitudinal
coupling with the two levels and does not affect the resonance transition.
For an isolated electronic state, the resonance transition occurs owing to the simultaneous
single-phonon--single-photon absorption, analogous to a two-photon absorption
process with different frequencies, such as bichromatic driving for ESR using
orthogonal electromagnetic waves.
\cite{Gromov00,Kalin04,Gyorgy22}
In quadrupole--quadrupole interaction systems, ordered moments can be probed using the
PA-MAR technique.
At the quadrupolar ordering transition, slight mixing between the two levels may induce a
transverse photon coupling, giving rise to an additional phonon--photon absorption process.
\par

%%%%%%%%%%%%%%%%%%%%%%%%%%%%%%%%%%%%%%
%%%%%%%%%%%%%%%%%%%%%%

On the basis of the Floquet theory,
\cite{Shirley65}
we formulate the PA-MAR for the two-level system coupled to two periodically oscillating fields:
a photon field and a phonon field with different frequencies.
\cite{Koga24b}
Although the longitudinal (diagonal) photon coupling does not contribute directly to the two-level
transition, it plays an important role in effectively reducing the excitation energy.
Consequently, the resonance transition becomes feasible despite the low frequency of the
phonon field because of the transverse (off-diagonal) phonon coupling.
Thus, the single-phonon absorption transition process can be potentially assisted by the
single-photon absorption.
We demonstrate how the PA-MAR transition probability depends on the propagation direction of
the acoustic wave, focusing on the $\Gamma_8$ quartet for the typical magnetic field
direction [110] in a cubic reference frame. 
\par

This paper is organized as follows.
Section~2 presents the formulation of the PA-MAR for a two-level system using Floquet
theory with a specific focus on the simultaneous single-phonon--single-photon absorption
transition process.
Section~3 discusses the time-averaged transition probability of PA-MAR for
the low-lying states in the $\Gamma_8$ quartet coupled to inplane strain fields induced by a
BAW.
The BAW propagation direction is changed while the magnetic field is aligned along the [110]
direction.
We also clarify how the induced AFQ ordered moment modifies the transition probability.
Section~4 provides a summary of the main findings.
Additional details are given in Appendices~A, B, and C.
Appendix~A introduces a pseudospin-3/2 representation for the $\Gamma_8$ quartet
considering the AFQ order in CeB$_6$.
Appendix~B discusses the case of a transverse BAW in comparison with that of a longitudinal
BAW considered in Sect.~3.2.
Appendix~C offers a concise overview of the relationship between BAW propagation
directions and displacement vectors within a cubic lattice.
\par

%%%%%%%%%%%%%%%%%%%%%%%%%%%%%%%%%%%%%%%%%%%%%%%%%%%%%%%%%%%%%%%%%%%%%%%%%%%%%%%%%%%%%%%%%%%%%%%%%%%
\section{Photon-assisted Magnetoacoustic Resonance}
%%%%%%%%%%%%%%%%%%%%%%%%%%%%%%%%%%%%%%%%%%%%%%%%%%%%%%%%%%%%%%%%%%%%%%%%%%%%%%%%%%%%%%%%%%%%%%%%%%%
%%%%%%%%%%%%%%%%%%%%%%%%%%%%%%%%%%%%%%%%%%%%%%%%%%%%%%%%%%%%%%%%%%%%%%%%%%%%%%%%%%%%%%%%%%%%%%%%%%%
\subsection{Two-level system coupled to photon and phonon fields}
%%%%%%%%%%%%%%%%%%%%%%%%%%%%%%%%%%%%%%%%%%%%%%%%%%%%%%%%%%%%%%%%%%%%%%%%%%%%%%%%%%%%%%%%%%%%%%%%%%%
Let us begin with the two-level system coupled to periodically oscillating photon
and phonon fields, as described by the following Hamiltonian:
\begin{align}
H (t) = \frac{1}{2} \left(
\begin{array}{cc}
- \varepsilon (t)  & h^* (t) \\
h(t) & \varepsilon (t)
\end{array}
\right),
\label{eqn:tildeH}
\end{align}
where
\begin{align}
& \varepsilon (t) = \varepsilon_0 + \Delta_L \cos ( \omega_\Delta t + \theta )
+ A_L \cos \omega_A t,
\label{eqn:ept} \\
& h (t) = \Delta_T \cos ( \omega_\Delta t + \theta ) + A_T \cos \omega_A t.
\label{eqn:ht}
\end{align}
The diagonal matrix elements are composed of the two-level splitting $\varepsilon_0$,
the longitudinal coupling to the photon field with amplitude $\Delta_L$  and frequency
$\omega_\Delta$, and the longitudinal coupling to the phonon field with $A_L$ and $\omega_A$.
Here, an initial phase shift $\theta$ is introduced to the photon field.
For off-diagonal matrix elements, $\Delta_T$ and $A_T$ represent the transverse coupling
constants of the photon and phonon fields, respectively.
The Schr\"{o}dinger equation with the periodically time-dependent Hamiltonian,
\begin{align}
i \hbar \frac{ \partial \psi ( t ) }{ \partial t } = H (t) \psi ( t ),
\end{align}
can be solved using Floquet theory, which is extended to the case of multifrequency.
By substituting $\psi (t ) = e^{ - i q t } \phi (t)$ into this equation and applying the Fourier expansion
\cite{Gyorgy22}
\begin{align}
H (t) = \sum_{n n'} H^{ [n, n'] } e^{ i  ( n \omega_\Delta + n' \omega_A ) t },~~
\phi (t) = \sum_{n n'} \phi^{ [n, n'] } e^{ i ( n \omega_\Delta + n' \omega_A ) t },
\label{eqn:Fexp}
\end{align}
we obtain the time-independent eigenvalue equation
\begin{align}
\sum_{m m'} \left\{ H^{ [ n - m, n' - m' ] } + ( n \omega_\Delta + n' \omega_A )
\delta_{nm} \delta_{n' m'} \right\} \phi^{ [ m, m' ] } = q \phi^{ [ n, n' ] },
\end{align}
and the solution is given for the quasi-energy $q$.
Here, $\hbar = 1$ for brevity.
This equation leads to the so-called Floquet Hamiltonian $H_F$,
\begin{align}
\langle \alpha n n' | H_F | \beta m m' \rangle = H_{\alpha \beta}^{ [ n - m, n' - m' ] }
+ ( n \omega_\Delta + n' \omega_A ) \delta_{\alpha \beta} \delta_{n m} \delta_{n' m'},
\end{align}
and $H_{\alpha \beta}^{ [ n - m, n' - m' ] } \equiv \langle \alpha | H^{ [ n - m, n' - m' ] } | \beta \rangle$
for the two levels denoted by $\alpha$ and $\beta$.
The infinite-dimensional matrix form of $H_F$ is constructed on the basis of the Floquet states
$| \alpha n n' \rangle = | \alpha \rangle \otimes | n \rangle \otimes | n' \rangle$ denoted
by the indices of level $\alpha$ and integer $n, n'$ ($ = 0, \pm 1, \pm 2, \cdots $) in
Eq.~(\ref{eqn:Fexp}).
The practical calculation is performed by solving
$H_F | q_\gamma \rangle = q_\gamma | q_\gamma \rangle$ with the $\gamma$th eigenvalue
$q_\gamma$, and the corresponding eigenvector $| q_\gamma \rangle$ using
the following $2 \times 2$ block matrices:
\begin{align}
& H^{ [0, 0] } = \frac{1}{2}
\left(
\begin{array}{cc}
- \varepsilon_0 & 0 \\
0 & \varepsilon_0
\end{array}
\right),~~
H^{ [ \pm 1, 0 ] } = \frac{ e^{\pm i \theta} }{4}
\left(
\begin{array}{cc}
- \Delta_L & \Delta_T^* \\
\Delta_T & \Delta_L
\end{array}
\right),
\nonumber \\
& H^{ [ 0, \pm 1 ] } = \frac{1}{4}
\left(
\begin{array}{cc}
- A_L & A_T^* \\
A_T & A_L
\end{array}
\right).
\end{align}
For the other block matrices, $H^{[ n - m, n' - m']} = {\bf 0}$ (zero matrix) with
$n - m = \pm 2, \pm 3, \cdots$ ($n' - m' \ne 0$), and it holds for the replacement
$( n, m ) \leftrightarrow ( n', m' )$.
To understand the structure of the matrix form of $H_F$, let us restrict the phonon sector to
$n' = 0$ and $m' = -1$ for $A_L = 0$ and $\Delta_T = 0$.
In the subspace of $\{ | \alpha, n, 0 \rangle, | \beta , m , -1 \rangle \}$
($n, m = \cdots, -1, 0, 1, \cdots$), the matrix form of $H_F^{\rm sub}$ is reduced to
\begin{align}
H_F^{\rm sub} =
\left(
\begin{array}{ccccc}
~ & ~ & \vdots & ~ & ~ \\
\ddots &  H^{[-1]} & {\bf 0} & {\bf 0} & ~ \\
~ & H_{-1}^{[0]} & H^{[-1]} & {\bf 0} & \cdots \\
~ & H^{[1]} & H_0^{[0]} & H^{[-1]} & ~ \\
\cdots & {\bf 0} & H^{[1]} & H_1^{[0]} & ~ \\
~ & {\bf 0} & {\bf 0} & H^{[1]} & \ddots \\
~ & ~ & \vdots & ~ & ~
\end{array}
\right),
\label{eqn:HFblock}
\end{align}
where
\begin{align}
& H_n^{[0]} = \left(
\begin{array}{cc}
- ( \varepsilon_0 / 2 ) - n \omega_\Delta & A_T^* / 4 \\
A_T / 4 & ( \varepsilon_0 / 2 ) - \omega_A - n \omega_\Delta
\end{array}
\right), \\
& H^{[ \pm 1 ]} = \frac{ e^{ \pm i \theta }}{4} \left(
\begin{array}{cc}
- \Delta_L & 0 \\
0 & \Delta_L
\end{array}
\right).
\end{align}
A similar matrix of $H_F^{\rm sub}$ is obtained for the other phonon sectors
$\{ | \alpha, n, k' \rangle, | \beta , m , k' \pm 1 \rangle \}$ ($k'$: integer) through the replacement
$- ( \varepsilon_0 / 2 ) \rightarrow - ( \varepsilon_0 / 2 ) + k' \omega_A$ and
$( \varepsilon_0 / 2 ) - \omega_A \rightarrow ( \varepsilon_0 / 2 ) + ( k' \pm 1 ) \omega_A$
in $H_n^{[0]}$.
Alternatively, the above argument also holds in the subspace of the photon sectors
$\{ | \alpha, k, n' \rangle, | \beta , k \pm 1 , m' \rangle \}$ ($k$: integer) for $A_T = 0$ and
$\Delta_L = 0$, where the parameters are replaced as $\omega_\Delta \leftrightarrow \omega_A$
and $A_T \rightarrow \Delta_T e^{ \pm i \theta }$ in $H_n^{[0]}$.
In $H^{[ \pm 1 ]}$, $\Delta_L \rightarrow A_L ( \theta = 0) $.  
\par

\subsection{Single-phonon--single-photon absorption transition}
To describe the resonance transition driven by multifrequency fields, it is convenient to use the
eigenstates of $H_F$ for both $A_T = 0$ and $\Delta_T = 0$,
\cite{Son09}
\begin{align}
& | \alpha n n' \rangle_L = \sum_{ k, k' = - \infty }^\infty e^{ i ( k - n ) \theta }
J_{ k - n } \left( \frac{ \Delta_L }{ 2 \omega_\Delta } \right)
J_{ k' - n' } \left( \frac{ A_L }{ 2 \omega_A } \right) | \alpha k k' \rangle,
\nonumber \\
& | \beta m m' \rangle_L = \sum_{ k, k' = - \infty }^\infty e^{ i ( k - m ) \theta }
J_{ k - m } \left( - \frac{ \Delta_L }{ 2 \omega_\Delta } \right)
J_{ k' - m' } \left( - \frac{ A_L }{ 2 \omega_A } \right) | \beta k k' \rangle,
\label{eqn:L-state}
\end{align}
with the $k$th Bessel function of the first kind $J_k$.
The subscript $L$ on the left-hand side indicates the longitudinal photon and phonon couplings. 
Here, we focus on the nearly degenerate Floquet states $| \alpha 0 0 \rangle_L$ and
$| \beta, - 1, -1 \rangle_L$ for the single $\pi$ photon and single longitudinal phonon absorption
processes under the condition
$- \varepsilon_0 / 2 \simeq ( \varepsilon_0 / 2 ) - \omega_\Delta - \omega_A$.
When both $A_T$ and $\Delta_T$ are finite, we treat these parameters as the perturbation,
\begin{align}
& \langle \alpha k  k' | H'_F | \beta l l' \rangle
= \frac{ A_T^* }{4} \delta_{k l} ( \delta_{k', l' + 1} + \delta_{k', l' - 1} )
\nonumber \\
&~~~~~~~~~~~~~~~~~~~~~~
+ \frac{ \Delta_T^* }{4} ( e^{ i \theta } \delta_{k, l + 1} + e^{ - i \theta } \delta_{k, l - 1} ) \delta_{k' l'}.
\end{align}
Using the formula for the Bessel function $J_n ( a + b ) = \sum_m J_m ( a ) J_{n - m} (b)$ and
$J_{n - 1} ( a ) + J_{ n + 1} ( a ) = ( 2 n / a ) J ( a )$, we calculate the perturbation terms with $A_T$
and those with $\Delta_T$ separately as
\begin{align}
& v_{k, k'}^A \equiv {}_L \langle \beta n n' | H'_F | \alpha m m' \rangle_L~( \Delta_T = 0 )
\nonumber \\
&~~~~~~
= \frac{ A_T }{2} e^{i k \theta} J_k \left( \frac{ \Delta_L }{ \omega_\Delta } \right)
\frac{ k' \omega_A }{ A_L } J_{k'} \left( \frac{ A_L }{ \omega_A } \right), \\
& v_{k, k'}^\Delta \equiv {}_L \langle \beta n n' | H'_F | \alpha m m' \rangle_L~( A_T = 0 )
\nonumber \\
&~~~~~~
= \frac{ \Delta_T }{2} e^{i k \theta} \frac{ k \omega_\Delta }{ \Delta_L }
J_k \left( \frac{ \Delta_L }{ \omega_\Delta } \right) J_{k'} \left( \frac{ A_L }{ \omega_A } \right),
\end{align}
where $k = n - m$ and $k' = n' - m'$.
In the subspace of the nearly degenerate Floquet states  $| \alpha 0 0 \rangle_L$ and
$| \beta, - 1, -1 \rangle_L$, the $2 \times 2$ matrix form of the effective Hamiltonian,
\begin{align}
H_{\rm eff} = \left(
\begin{array}{cc}
\ds{ - \frac{ \varepsilon_0 }{2} } + \delta & v_{ - 1, -1 }^* \\
v_{ - 1, -1 } & \ds{ \frac{ \varepsilon_0 }{2} } - \delta - \omega_A - \omega_\Delta
\end{array}
\right),
\end{align}
describes the transition at $\varepsilon_0 \simeq \omega_A + \omega_\Delta$.
Here, the off-diagonal term $v_{-1, -1}$ is related to the peak broadening of the transition
probability,
\begin{align}
v_{ - 1, -1 } = v_{ -1, -1 }^A + v_{ - 1, -1 }^\Delta
\simeq - \frac{ e^{ - i \theta } }{8} \left( \frac{ A_T \Delta_L }{ \omega_\Delta }
+ \frac{ \Delta_T A_L }{ \omega_A } \right),
\label{eqn:v-1-1}
\end{align}
for the weak coupling condition $| A_L | / \omega_A, | \Delta_L | / \omega_\Delta \ll 1$, where
the approximation $J_k ( a ) = ( - 1 )^k J_{ - k } ( a ) \simeq ( a / 2 )^k / k !$ ($k \geq 0$) is used.
Like the Bloch--Siegert shift for the optically driven resonance transition, the energy
shift $\delta$ is calculated using the second-order perturbation with respect to
$v_{k, \pm 1}^A$ ($\propto A_T$) and $v_{\pm 1, k}^\Delta$ ($\propto \Delta_T$) as
\begin{align}
& \delta =
- \sum_{ {}_{ k \ne - 1 }^{ k  = - \infty } }^\infty
\left[ \frac{ | v_{k, -1}^A |^2 }{ \varepsilon_0 - \omega_A + k \omega_\Delta }
+ \frac{ | v_{k, 1}^A |^2 }{ \varepsilon_0 + \omega_A + k \omega_\Delta } \right.
\nonumber \\
&~~~~~~~~~~~~~~\left.
+ \frac{ | v_{-1, k}^\Delta |^2 }{ \varepsilon_0 - \omega_\Delta + k \omega_A }
+ \frac{ | v_{1, k}^\Delta |^2 }{ \varepsilon_0 + \omega_\Delta + k \omega_A } \right].
\end{align}
For weak coupling, the leading term of $\delta$ is represented by the sum of only the $k = 0$
terms,
\begin{align}
\delta \simeq - \frac{ \varepsilon_0 | A_T |^2 }{ 8 ( \varepsilon_0^2 - \omega_A^2 ) }
- \frac{ \varepsilon_0 | \Delta_T |^2 }{ 8 ( \varepsilon_0^2 - \omega_\Delta^2 ) }.
\end{align}
For the transition from $| \alpha \rangle$ to $| \beta \rangle$ in the two states, the time-averaged
probability can be derived from the eigenvalues of $H_{\rm eff}$ as
\cite{Son09}
\begin{align}
\bar{P}_{\alpha \rightarrow \beta} =
\frac{1}{2} \frac{ | v_{-1, -1} |^2 }
{ | v_{-1, -1} |^2 + ( \omega_\Delta + \omega_A  - \varepsilon_0 + 2 \delta )^2 / 4}.
\end{align}
In particular, we focus on
$\bar{P}_{\alpha \rightarrow \beta}$ at $\varepsilon_0 = \omega_\Delta + \omega_{A}$
for the simultaneous single-phonon--single-photon absorption transition,
\begin{align}
\bar{P}_{\alpha \rightarrow \beta}^{(1,1)} =
\frac{1}{2} \frac{1}{ 1 + \gamma^2 }~~\left( \gamma \equiv \frac{ | \delta | }{ | v_{-1, -1} | } \right),
\label{eqn:TP}
\end{align}
except for $A_T = \Delta_T = 0$, and the ratio $\gamma$ is obtained as
\begin{align}
& \gamma = \frac{ | A_T | ( 1 + \bar{\omega} )
\left( \ds{ \frac{ \bar{\omega} }{ 1 + 2 \bar{\omega} }
+ \frac{1}{ 2 + \bar{\omega} } \frac{ | \Delta_T |^2 }{ | A_T |^2 } } \right) }
{ | \Delta_L | \left| \bar{\omega} + \ds{ \frac{ \Delta_T A_L }{ \Delta_L A_T } } \right| }
~~\left( \bar{\omega} \equiv \frac{ \omega_A }{ \omega_\Delta } \right)
\nonumber \\
&~~
\simeq \left\{
\begin{array}{ll}
| A_T | / | \Delta_L | & ( | \Delta_T | / | A_T | \ll \bar{\omega} \ll 1 ) \\
| \Delta_T | / ( 2 | A_L | ) & ( \bar{\omega} \ll | \Delta_T | / | A_T | \sim | \Delta_L | / | A_L | ).
\end{array}
\right.
\label{eqn:gamma}
\end{align}

%%%%%%%%%%%%%%%%%%%%%%%%%%%%%%%%%%%%%%%%%%%%%%%%%%%%%%%%%%%%%%%%%%%%%%%%%%%%%%%%%%%%%%%%%%%%%%%%%%%
\section{Acoustically Driven Resonance Transition}
%%%%%%%%%%%%%%%%%%%%%%%%%%%%%%%%%%%%%%%%%%%%%%%%%%%%%%%%%%%%%%%%%%%%%%%%%%%%%%%%%%%%%%%%%%%%%%%%%%%
\subsection{Inplane quadrupole--strain interaction}
For the quadrupole operators constructed by the spin $S = 3/2$ operators, we consider three
components, namely,
\begin{align}
& O_u = \frac{1}{ \sqrt{3} } ( 2 S_z^2 - S_x^2 - S_y^2 ),~~O_v = S_x^2 - S_y^2,
\nonumber \\
& O_{xy} = S_x S_y  + S_y S_x,
\end{align}
for the inplane ($xy$-plane) QS interaction.
The lattice deformations in the $xy$-plane are described by the three components of the strain
tensors $\varepsilon_{xx} = \partial u_x / \partial x$, $\varepsilon_{yy} = \partial u_y / \partial y$,
and $\varepsilon_{xy} = [ ( \partial u_y / \partial x ) + ( \partial u_x / \partial y ) ] / 2$, where
$\bu = ( u_x, u_y, 0 )$ is the displacement vector.
The $u$- and $v$-type quadrupoles ($O_u$ and $O_v$) are coupled to the corresponding strain
tensors defined as
$\varepsilon_u \equiv ( 2 \varepsilon_{zz} - \varepsilon_{xx} - \varepsilon_{yy} ) / \sqrt{3}$ and
$\varepsilon_v \equiv \varepsilon_{xx} - \varepsilon_{yy}$, respectively, and the $xy$ type is
coupled to $\varepsilon_{xy}$.
The QS interaction Hamiltonian is given by
\cite{Dohm75,Luthi05}
\begin{align}
H_\varepsilon = g_a O_u \varepsilon_u + g_b O_v \varepsilon_v
+ g_c O_{xy} \varepsilon_{xy},
\label{eqn:Hep}
\end{align}
with the three independent coupling constants $g_a$, $g_b$, and $g_c$
($g_a = g_b$ is satisfied for $O_h$).
For the magnetic field $\bH \parallel Z \parallel [110]$, we transform the quadrupole operators
in Eq.~(\ref{eqn:Hep}) to those redefined in the $(XYZ)$ reference frame using the unitary matrix
$U$ in Eq.~(\ref{eqn:US}).
The unitary transformed $O_k$ ($k = u, v, xy$) are given by the linear combinations of $O_K$
($K = U, V, YZ$),
\begin{align}
& U^\dagger O_u U = - \frac{1}{2} O_U + \frac{ \sqrt{3} }{2} O_V,~~
U^\dagger O_v U = O_{YZ},
\nonumber \\
& U^\dagger O_{xy} U = \frac{ \sqrt{3} }{2}  O_U + \frac{1}{2} O_V.
\label{eqn:OkOK}
\end{align}
For the transformed QS interaction Hamiltonian,
\begin{align}
\tilde{H}_\varepsilon = U^\dagger H_\varepsilon U = \sum_K A_K O_K,
\end{align}
the strain-dependent coupling coefficients are written as
\begin{align}
& A_U = - \frac{1}{2} g_a \varepsilon_u + \frac{ \sqrt{3} }{2} g_c \varepsilon_{xy},~~
A_V = \frac{ \sqrt{3} }{2} g_a \varepsilon_u + \frac{1}{2} g_c  \varepsilon_{xy},
\nonumber \\
& A_{YZ} = g_b \varepsilon_v.
\label{eqn:AK}
\end{align}
For the basis of $\{ | 3/2 \rangle, | 1/2 \rangle, | - 1/2 \rangle, | - 3/2 \rangle \}$ with the quantization
axis $Z$, the matrices of the corresponding quadrupole operators are given by
\cite{Koga24b}
\begin{align}
& \frac{ O_U }{ \sqrt{3} } = \left(
\begin{array}{cccc}
1 & 0 & 0 & 0 \\
0 & -1 & 0 & 0 \\
0 & 0 & -1 & 0 \\
0 & 0 & 0 & 1
\end{array}
\right),~~
\frac{ O_V }{ \sqrt{3} } = \left(
\begin{array}{cccc}
0 & 0 & 1 & 0 \\
0 & 0 & 0 & 1 \\
1 & 0 & 0 & 0 \\
0 & 1 & 0 & 0
\end{array}
\right),
\nonumber \\
& \frac{ O_{YZ} }{ \sqrt{3} } = \left(
\begin{array}{cccc}
0 & - i & 0 & 0 \\
i & 0 & 0 & 0 \\
0 & 0 & 0 & i \\
0 & 0 & - i & 0
\end{array}
\right),
\end{align}
respectively.
\par

In the subsequent discussion, we concentrate on the two lowest-lying states of the crystal field
quartet that participate in an antiferroquadrupolar ordering system exemplified by CeB$_6$, as
described in Appendix~A.
The coupling coefficients with the phonon fields in Eqs.~(\ref{eqn:ept}) and (\ref{eqn:ht}) are
directly derived from the matrix elements of $\tilde{H}_\varepsilon$.
In the subspace of $\{ | \psi_{\mu, 1} \rangle, | \psi_{\mu, 2} \rangle \}$, which are the eigenstates of
the mean field quadrupole--quadrupole interaction Hamiltonian $H_{\rm local}^{(\pm)}$ for an
induced quadrupole moment $\mu_{\rm af}$ in Eq.~(\ref{eqn:Hlocal}), we obtain
\begin{align}
& A_L = \langle \psi_{\mu, 2} | \tilde{H}_\varepsilon | \psi_{\mu, 2} \rangle
- \langle \psi_{\mu, 1} | \tilde{H}_\varepsilon | \psi_{\mu, 1} \rangle
\nonumber \\
&~~~~
= 2 \sqrt{3} ( - A_U \cos \theta_\lambda + A_V \sin \theta_\lambda ) \cos 2 \theta_\mu, \\
& A_T = 2 \langle \psi_{\mu, 2} | \tilde{H}_\varepsilon | \psi_{\mu, 1} \rangle
= ( \pm \tan 2 \theta_\mu ) A_L + i 2 \sqrt{3} A_{YZ},
\label{eqn:AT}
\end{align}
where $\cos \theta_\lambda = ( 16 + 3 \lambda ) / F_\lambda$ and
$\sin \theta_\lambda = 15 \sqrt{3} \lambda / F_\lambda$, using $F_\lambda$, as defined in
Eq.~(\ref{eqn:cFlambda}).
The parameter $\lambda$ quantifies the anisotropy of the Zeeman effect for the $\Gamma_8$
quartet.
In a similar way, the $\pi$ photon field has the longitudinal coupling of
$\Delta_L \propto \langle \psi_{\mu, 2} | \tilde{J}_Z | \psi_{\mu, 2} \rangle
- \langle \psi_{\mu, 1} | \tilde{J}_Z | \psi_{\mu, 1} \rangle$, where these matrix elements are given
in Eq.~(\ref{eqn:JZ}), and $\tilde{J}_Z$ is the effective dipole operator described in Appendix~A.
The transverse photon coupling is generated by the induced quadrupole moment $\mu_{\rm af}$
as $\Delta_T = ( \pm \tan 2 \theta_\mu ) \Delta_L $.
Rewriting $A_T \rightarrow | A_T | e^{ i \theta_T}$, we obtain
$\cos \theta_T = \pm \tan 2 \theta_\mu~( A_L / | A_T | )$.
Because the finite moment $\mu_{\rm af}$ also causes an additional energy splitting,
$\gamma$ in Eq.~(\ref{eqn:TP}) is modified as
$\gamma_\mu = | - \delta + ( \delta \varepsilon_\mu / 2 ) | / | v_{-1, -1} |$ for
$| A_T | / \omega_\Delta, | \Delta_T | / \omega_A \ll 1$, where
$\delta \varepsilon_\mu = ( \omega_\Delta + \omega_A )
[ \sqrt{ 1 + ( \bar{\mu}^2 / E_0^2 ) } - 1 ]$ ($\propto \bar{\mu}^2 \ll 1$).
As defined in Eq.~(\ref{eqn:E0mu}), $E_0$ is related to the Zeeman splitting
($\omega_\Delta + \omega_A = 2 E_0 B$), and $\bar{\mu}$ represents the value of $\mu_{\rm af}$
multiplied by $D_Q / B$.
For the transition probability
$\bar{P}_{1 \rightarrow 2}^{(1,1)} = (1/2) ( 1 + \gamma_\mu^2 )^{-1}$ between
$| \psi_{\mu,1} \rangle$ and $| \psi_{\mu, 2} \rangle$, $\gamma_\mu$ is calculated as
\begin{align}
& \gamma_\mu^2 = \frac{ | A_T |^2 }{ \Delta_L^2 }
\frac{ ( 1 + \bar{\omega} )^2 }{ \bar{\omega}^2 + ( 1 + 2 \bar{\omega} ) \cos^2 \theta_T }
\nonumber \\
&~~~~~~\times
\left[ \frac{ \bar{\omega} }{ 1 + 2 \bar{\omega} }
+ \frac{ \tan^2 2 \theta_\mu }{ 2 + \bar{\omega} } \frac{ \Delta_L^2 }{ | A_T |^2 }
+ \frac{ 4 \omega_A \delta \varepsilon_\mu }{ ( 1 + \bar{\omega} ) | A_T |^2 } \right]^2,
\label{eqn:gamma2}
\end{align}
where $\delta \varepsilon_\mu / \omega_\Delta \ll 1$ is assumed, and
$\bar{\omega} \equiv \omega_A / \omega_\Delta$.
The second term in the brackets $[ \cdots ]$ is negligibly small compared with the last term for the
weak coupling condition $| \Delta_L | / \omega_\Delta \ll 1$.
As a result, Eq.~(\ref{eqn:gamma2}) is simplified as
\begin{align}
\gamma_\mu^2 = \frac{ | A_T |^2 }{ \Delta_L^2 }
\frac{ \bar{\omega}^2 ( 1 + \bar{\omega} )^2 }{ \bar{\omega}^2 + ( 1 + 2 \bar{\omega} )
\cos^2 \theta_T } \left( \frac{1}{ 1 + 2 \bar{\omega} }
+ \frac{ \bar{\mu}_{\rm eff}^2 }{ 1 + \bar{\omega} } \frac{ \Delta_L^2 }{ | A_T |^2} \right)^2,
\label{eqn:gamma2ef}
\end{align}
where the effective induced moment $\bar{\mu}_{\rm eff}$ is defined as
\begin{align}
& \bar{\mu}_{\rm eff}^2 \equiv 4 ( \omega_\Delta / \Delta_L )^2
( \delta \varepsilon_\mu / \omega_\Delta )
\nonumber \\
&~~~~~~
\simeq 2 ( 1 + \bar{\omega} ) ( \omega_\Delta / \Delta_L )^2 ( \bar{\mu} / E_0 )^2~~
(\bar{\mu} \simeq 0).
\end{align}
Note that the calculated $\gamma_\mu$ is independent of the $\pm$ signs of the staggered 
moment as defined in Eq.~(\ref{eqn:Hlocal}), i.e., $(\pm \tan 2 \theta_\mu)$ in Eq.~(\ref{eqn:AT}).
\par

\subsection{Quadrupole transition depending on propagation directions of a bulk acoustic wave}
We consider that the local pseudo-quartet is coupled to the strain fields induced by a BAW
propagating in the $x'$ direction, denoted by angle $\varphi$, which is measured relative to the
direction of the crystallographic $x$-axis.
The $x'$ and  $y'$ coordinates of BAW are transformed as
\begin{align}
\left(
\begin{array}{c}
x' \\
y'
\end{array}
\right)
= \left(
\begin{array}{cc}
\cos \varphi & \sin \varphi \\
- \sin \varphi & \cos \varphi
\end{array}
\right)
\left(
\begin{array}{c}
x \\
y
\end{array}
\right).
\label{eqn:rotation}
\end{align}
In this study, we focus on a longitudinal BAW considering the displacement of the vibration
parallel to the propagation direction ($x'$).
The three strain tensor components in Eq.~(\ref{eqn:AK}) are replaced as
\begin{align}
\varepsilon_u  \rightarrow - \frac{ \varepsilon_{x' x'} }{ \sqrt{3} },~~
\varepsilon_v \rightarrow \varepsilon_{x' x'} \cos 2 \varphi,~~
\varepsilon_{xy} \rightarrow \frac{ \varepsilon_{x' x'} }{2} \sin 2 \varphi.
\end{align}
As indicated in Appendix~C, the displacement vector $\bu$ is not parallel to the propagation
direction in the $xy$ plane of a cubic crystal, except for the principal $[100]$ and diagonal $[110]$
axes.
As discussed subsequently, our focus is on the transition probabilities driven by the BAW
propagating along the $x'$-axis in the vicinity of [110] ($\varphi = \pi / 4$) and [$\bar{1}$10]
($\varphi = 3 \pi / 4$).
Consequently, we represent $\bu$ by the single component $u_{x'}$.
Using Eq.~(\ref{eqn:AK}), $| A_T |^2$ in Eq.~(\ref{eqn:AT}) is given as a function of $\varphi$,
\begin{align}
| A_T |^2 = \left( \frac{ \bar{\mu} }{ F_\lambda E_{ \bar{\mu} } } \right)^2 a_L ( \varphi )
+ a_{YZ} ( \varphi ),
\label{eqn:ATf}
\end{align}
where
\begin{align}
& a_L ( \varphi ) = 4 \left[ 8 ( 1 + 3 \lambda ) g_a + 3 ( 4 - 3 \lambda ) g_c \sin 2 \varphi \right]^2
\langle \varepsilon_{x' x'}^2 \rangle,
\label{eqn:fR} \\
& a_{YZ} ( \varphi ) = 12 ( g_b \cos 2 \varphi )^2 \langle \varepsilon_{x' x'}^2 \rangle.
\label{eqn:fI}
\end{align}
The square of the strain amplitude
$\langle \varepsilon_{x' x'}^2 \rangle$ is represented by the average with respect to the spatial
distribution of magnetic ions.
In Eq.~(\ref{eqn:ATf}), $a_{YZ} ( \varphi )$ approaches zero as $\varphi \rightarrow \pi / 4$
($x' \parallel H \parallel [110]$) and $\varphi \rightarrow 3 \pi / 4$ ($x' \bot H$).
For these values of $\varphi$, we obtain $| A_T |^2 / A_L^2 = ( \bar{\mu} / E_0 )^2$, and
Eq.~(\ref{eqn:gamma2ef}) is reduced to
$\gamma_\mu^2 = \bar{\mu}^2 ( 2 \omega_\Delta \omega_A )^2 / ( E_0 \Delta_L A_L )^2$ under
the condition $| \Delta_L | / \omega_\Delta \ll 1$.
For other values of $\varphi$, the transition probability is not significantly affected by a finite
$\bar{\mu}$.
\par

%%%%%%%%%%%%%%%%%%%%%%%%%%%%%%%%%%%%%%
\begin{figure}
\begin{center}
\includegraphics[width=7cm,clip]{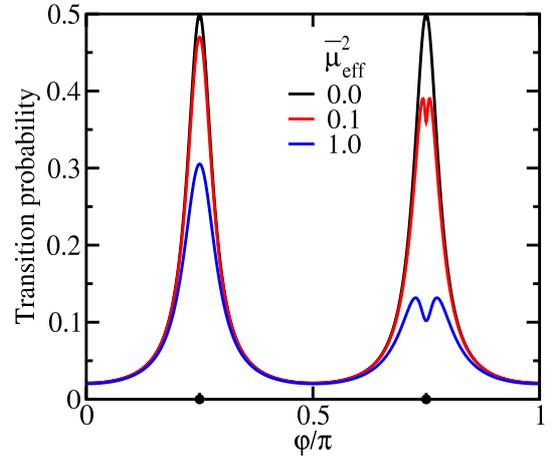}
\end{center}
\caption{
(Color online)
Transition probability $\bar{P}_{1 \rightarrow 2}^{(1,1)}$ plotted as a function of the BAW
propagation direction $\varphi$ for various values of the effective induced moment
$\bar{\mu}_{\rm eff}^2 = 0.0$, $0.1$, and $1.0$.
At the normal phase ($\bar{\mu}_{\rm eff} = 0$), $\bar{P}_{1 \rightarrow 2}^{(1,1)}$ vanishes
precisely at $\varphi / \pi = 1/4$ and $3/4$.
Here, the magnetic field $\bH$ is parallel to the [110] direction ($\varphi = \pi /4$).
}
\label{fig:2}
\end{figure}
%%%%%%%%%%%%%%%%%%%%%%

In Fig.~\ref{fig:2}, we show the $\varphi$ dependence of the transition probability for various
values of $\bar{\mu}_{\rm eff}^2$ ($= 10^4 \bar{\mu}^2$), considering the symmetric QS coupling
case ($g_a = g_b = g_c \equiv g$).
The other parameters are fixed as $\lambda = 4/9$, $\bar{\omega} = 0.1$, and
$g^2 \langle \varepsilon_{x' x'}^2 \rangle / \Delta^2 = 1$.
Here, the coupling strength $\Delta$ of the photon field is related to $\Delta_L$ as
$\Delta_L = 2 \Delta E_0^2 / E_{ \bar{\mu} }
= ( 2 \Delta / 3 ) ( 1 + 9 \bar{\mu}^2 )^{ - 1/ 2 }$.
As discussed above, the two peaks appear at $\varphi = \pi / 4$ and $\varphi = 3 \pi / 4$, and
their intensities show a marked reduction with the increase in $\bar{\mu}^2$.
The difference between the $\varphi$ dependence in $0 < \varphi < \pi / 2$ and that in
$ \pi / 2 < \varphi < \pi$ is due to the contribution from the term including $g_a g_c \sin 2 \varphi$
in Eq.~(\ref{eqn:fR}).
At the peak for $\varphi = 3 \pi / 4$, the appearance of the dip signifies the competition
between the QS coupling and quadrupole ordering effects.
When the former effect is diminished by reducing the coupling $g$, a similar dip also emerges
at $\varphi = \pi / 4$ and deepens with the enhancement of $\bar{\mu}^2$.
\par

This result demonstrates that the quadrupole components $O_U$ and $O_{V}$ in the longitudinal
coupling are involved in the two-level transition when the $O_{ZX}$ ordered moment $\bar{\mu}$
emerges.
In the transverse coupling in Eq.~(\ref{eqn:AT}), $A_T$ contains only $A_{YZ}$, and the
single component $O_{YZ}$ contributes to the transition in the normal phase.
For a finite $\bar{\mu}$, the diagonal matrix element $A_L$ ($A_U$ and $A_V$)  as well as
$A_{ZX}$ appears in $A_T$.
Because of the two-level mixing, the abrupt change in transition probability is attributed to the
quadrupole transition forbidden in the normal phase, which provides the evidence of the
off-diagonal order such as $O_{ZX}$ considered in the present study.
\par

Finally, we mention that the single-photon absorption transition occurs for a finite $\bar{\mu}$
when $\omega_\Delta$ is adjusted to the energy difference between $| \psi_{\mu, 1} \rangle$ and
$| \psi_{\mu, 2} \rangle$.
The photon-assisted single-phonon absorption transition can be observed as the sideband
resonance.
\par

%%%%%%%%%%%%%%%%%%%%%%%%%%%%%%%%%%%%%%%%%%%%%%%%%%%%%%%%%%%%%%%%%%%%%%%%%%%%%%%%
\section{Conclusion}
%%%%%%%%%%%%%%%%%%%%%%%%%%%%%%%%%%%%%%%%%%%%%%%%%%%%%%%%%%%%%%%%%%%%%%%%%%%%%%%%
We presented a theory of PA-MAR as a novel magnetoacoustic resonance technique, combining
a $\pi$-photon field with acoustically induced strain fields to probe quadrupole characteristics of a
crystal field quartet.
Owing to an upper limit on acoustic wave frequencies, this method effectively realizes
quadrupole resonance transitions for large excitation energy gaps under strong magnetic fields,
which are inevitably required for precise measurements of the $\Gamma_8$ quartet quadrupoles
in CeB$_6$.
On the basis of the Floquet theory, we developed a block matrix form of the Floquet Hamiltonian
for a two-level system coupled to both periodically oscillating photon and phonon fields.
In particular, we focused on the time-averaged PA-MAR transition probability under a weak
coupling condition.
\par

In this study, we investigated the quadrupole transition between the two lowest $\Gamma_8$
levels in a magnetic field along the [110] axis.
The transition depends on a longitudinal BAW propagation direction $\varphi$ measured from
the [100] axis.
The photon and phonon frequencies are fixed at the values tuned to satisfy the PA-MAR
condition in the normal phase.
The transition probability exhibits distinct peaks at $\varphi \rightarrow \pi / 4$ ([110]) and
$\varphi \rightarrow 3 \pi / 4$ ([$\bar{1}$10]), reflecting the fourfold symmetry of
the quadrupole.
These peaks show an abrupt reduction when an ordered moment is induced by the off-diagonal
AFQ order within the two levels.
This behavior indicates that the quadrupole components contributing to the longitudinal
coupling, which are inactive in the normal phase, participate in the
PA-MAR transition process.
Simultaneously, the transverse photon coupling is also generated.
\par

Our proposal of PA-MAR is directly applicable to other pseudospin-3/2 quartets, such as the Re
$5d^1$ electron state realized in the cubic Mott insulator Ba$_2$MgReO$_6$.
\cite{Marjerrison16,Hirai19,Hirai20}
In the $5d^1$ configuration, $\Gamma_8$ can be a ground state originating from the strong
spin--orbit coupling for the threefold degenerate $t_{2g}$ orbital.
Ba$_2$MgReO$_6$ exhibits successive quadrupolar and magnetic ordering transitions
at 33~K and 18~K, respectively.
In the quadrupolar ordering phase, the coexistence of antiferroically arranged $O_v$ and
ferroically arranged $O_u$ moments was observed.
The origin of this exotic phenomenon has been attributed to the quadrupole components
possessed by the $\Gamma_8$ quartet.
\cite{Hirai20}
Therefore, it will be intriguing to confirm whether the experimentally observed quadrupole
components are predominantly governed by $\Gamma_8$ and clarify the distinct quadrupole
characteristics between the $d$-electron and $f$-electron systems.
This distinction is nontrivial because crystal field effects generally dominate over spin--orbit
coupling in localized $d$-electron states.
\cite{Kubo23}
\par

%%%%%%%%%%%%%%%%%%%%%%%%%%%%%%%%%%%%%%%%%%%%%%%%
%%%%%%%%%%%%%%%%%%%%%%%%%%%%%%%%%%%%%%%%%%%%%%%%%%%
%\acknowledgments
%%%%%%%%%%%%%%%%%%%%%%%%%%%%%%%%%%%%%%%%%%%%%%%%%%%%%%%%%%%%%%%%%%%%%%%%%%%%%%%%%%%%%%%%%%%%%%%%%%%
%This work was supported by JSPS KAKENHI Grant Number 21K03466.
\bigskip
{\footnotesize
{\bf Acknowledgments}~~
The authors are deeply grateful to T. Yanagisawa for his insightful comments on the potential of
optically assisted magnetoacoustic resonance measurements.
This work was supported by JSPS KAKENHI Grant Number 21K03466.
}

%%%%%%%%%%%%%%%%%%%%%%%%%%%%%%%%%%%%%%%%%%%%%%%%%%%%%%%%%%%%%%%%%%%%%%%%%%%%%%%%
\appendix
%%%%%%%%%%%%%%%%%%%%%%%%%%%%%%%%%%%%%%%%%%%%%%%%%%%%%%%%%%%%%%%%%%%%%%%%%%%%%%%%
%%%%%%%%%%%%%%%%%%%%%%%%%%%%%%%%%%%%%%%%%%%%%%%%%%%%%%%%%%%%%%%%%%%%%%%%%%%%%%%%
\section{Effect of Staggered Moment of Antiferroquadrupolar Order}
%%%%%%%%%%%%%%%%%%%%%%%%%%%%%%%%%%%%%%%%%%%%%%%%%%%%%%%%%%%%%%%%%%%%%%%%%%%%%%%%
We here consider the two lowest-lying states of the crystal field quartet involved in a quadrupolar
order.
One good example is the AFQ order in Ce$B_6$ for the applied magnetic field
$\bH \parallel [110]$.
The linear combination of the $\Gamma_5$-type quadrupole components
$O_{yz}$, $O_{zx}$, and $O_{xy}$ ($O_{\mu \nu} = J_\mu J_\nu + J_\nu J_\mu$)
is considered a prime candidate for the order parameter.
\cite{Shiina97,Sakai97,Shiina98,Matsumura12,Portnichenko20}
Here, $J_\mu$ is a dipole component ($\mu = x, y, z$).
The most dominant quadrupole coupling between the two states is associated with
$O_{yz} + O_{zx}$ for $\bH \parallel [110]$ in the cubic reference frame with the $x$, $y$, and $z$
coordinates.
\cite{Shiina97,Sakai97,Shiina98}
When the $\Gamma_8$ quartet in the Ce $f^1$ configuration is represented by a pseudospin
$S = 3/2$, the dipole and quadrupole operators are given by $4 \times 4$ matrices for the
basis of the eigenstates of $S_z$, $\{ | 3/2 \rangle, | 1/2 \rangle, | - 1/2 \rangle, | - 3/2 \rangle \}$.
To calculate the quadrupole transition matrix elements, it is convenient to use a different reference
frame where the in-plane field direction is chosen as the quantization axis $Z$, i.e.,
$\bH = H \be_Z$ with $\be_Z = (1, 1, 0) / \sqrt{2}$.
The other orthogonal unit vectors are defined as $\be_X = (0, 0 ,1)$ and
$\be_Y = (1, - 1, 0) / \sqrt{2}$.
For a finite magnetic field, it is convenient to transform the dipole and quadrupole operators
defined in the cubic ($xyz$) reference frame to those redefined in the new ($XYZ$) reference
frame using the unitary matrix,
\cite{Koga24b}
\begin{align}
U = \frac{1}{ 2 \sqrt{2} } \left(
\begin{array}{cccc}
\chi^{-3} & \sqrt{3} \chi^{-3} & \sqrt{3} \chi^{-3} & \chi^{-3} \\
\sqrt{3} \chi^{-1} & \chi^{-1} & - \chi^{-1} & - \sqrt{3} \chi^{-1} \\
\sqrt{3} \chi & - \chi & - \chi & \sqrt{3} \chi \\
\chi^3 & - \sqrt{3} \chi^3 & \sqrt{3} \chi^3 & - \chi^3
\end{array}
\right),
\label{eqn:US}
\end{align}
where $\chi = e^{i \pi / 8}$.
This leads to $S_\mu = U^\dagger ( \be_\mu \cdot \bS ) U$
($\mu = X, Y, Z$) for the dipole operator $\bS = (S_x, S_y, S_z)$.
The Zeeman splitting for $\bH \parallel Z$ is described by the Hamiltonian
\begin{align}
H_{\rm Z} = - g_J \mu_{\rm B} H
\sum_{i = 1, 2} E_{{\rm Z}, i} | \psi_{ {\rm Z},i } \rangle \langle \psi_{ {\rm Z}, i } |,
\label{eqn:HZ}
\end{align}
for the two lowest-lying states:
\cite{Koga24b}
\begin{align}
| \psi_{ {\rm Z}, 1 } \rangle = c_+ | 3/2 \rangle + c_- | - 1/2 \rangle,~~
| \psi_{ {\rm Z}, 2 } \rangle = - c_- | - 3/2 \rangle + c_+ | 1/2 \rangle,
\end{align}
where $| m \rangle$ ($m = \pm 1/2, \pm 3/2$) represents the eigenstates of $S_Z$, and
\begin{align}
c_\pm = \pm \sqrt{ \frac{1}{2} \left( 1 \pm \frac{ 16 + 3 \lambda }{F_\lambda} \right) },~~
F_\lambda = \sqrt{ ( 16 + 3 \lambda )^2 + ( 15 \sqrt{3} \lambda )^2 }.
\label{eqn:cFlambda}
\end{align}
In Eq.~(\ref{eqn:HZ}), $\mu_{\rm B}$ is the Bohr magneton and $g_J$ is Land{\' e}'s $g$ factor.
For the eigenenergies, $F_\lambda$ is related to
$E_{ {\rm Z}, 1 } = ( 8 - 6 \lambda + F_\lambda ) / 16$ and
$E_{ {\rm Z}, 2 } = ( - 8 + 6 \lambda + F_\lambda ) / 16$.
Note that the $Z$ component of the effective dipole operator in the two states is given by
$\tilde{J}_Z = - \sum_{i = 1, 2} E_{ {\rm Z}, i } | \psi_{ {\rm Z},i } \rangle \langle \psi_{ {\rm Z}, i } |$.
The parameter $\lambda$ represents an octupole effect that induces an anisotropic Zeeman
effect.
In the case of an $f^1$ $\Gamma_8$ quartet, $\lambda$ is equal to $4/9$.
\par

The quadrupole operator $O_{yz} + O_{zx}$ corresponds to $O_{ZX}$ in the ($XYZ$) reference frame, and the latter generates an off-diagonal coupling between the two states as
$\mu = | \psi_{ {\rm Z}, 2 } \rangle \langle  \psi_{ {\rm Z}, 1 } |
+  | \psi_{ {\rm Z}, 1 } \rangle \langle  \psi_{ {\rm Z}, 2 } |$.
For the AFQ ordering, we define a staggered moment $\mu_{\rm af}$ for the
intersite $O_{ZX}$ quadrupole--quadrupole interaction in the mean-field Hamiltonian for the
localized electron states,
\cite{Koga24b}
\begin{align}
H_{\rm local}^{ ( \pm ) } = D_Q ( \mp \mu_{\rm af} )~\mu + H_{\rm Z},
\label{eqn:Hlocal}
\end{align}
where $D_Q$ ($ > 0$) represents the strength of the interaction, and the signs $\pm$ of the first
term are chosen for the two sublattice sites.
The eigenvalue equation
$H_{\rm  local}^{ ( \pm ) } | \psi_{\mu, j} \rangle = B E_{\mu, j} | \psi_{\mu, j} \rangle$
($B \equiv g_J \mu_{\rm B} H$) is solved as
$| \psi_{\mu, j} \rangle = \sum_{i = 1, 2} c_{ij} | \psi_{ {\rm Z}, i } \rangle$ with
$E_{\mu, 1} = - ( F_\lambda / 16 ) - E_{ \bar{\mu} }$ and
$E_{\mu, 2} = - ( F_\lambda / 16 ) + E_{ \bar{\mu} }$,
where $E_{ \bar{\mu} } = \sqrt{ E_0^2 + \bar{\mu}^2 }$, and
\begin{align}
E_0 \equiv \frac{ 4 - 3 \lambda }{8},~~\bar{\mu} \equiv \frac{ D_Q \mu_{\rm af} }{B}.
\label{eqn:E0mu}
\end{align}
The coefficients $\{ c_{ij} \}$ are given by $c_{11} = c_{22} = \cos \theta_\mu$ and
$c_{21} = - c_{12} = \pm \sin \theta_\mu$, which are related to $E_{ \bar{\mu} }$ as
$\cos 2 \theta_\mu = E_0 / E_{ \bar{\mu} }$ and
$\sin 2 \theta_\mu = \bar{\mu} / E_{ \bar{\mu} }$.
Using these parameters, the induced quadrupole moment $\mu_{\rm af}$ modifies the matrix
elements of $\tilde{J_Z}$ as
\begin{align}
& \langle \psi_{\mu, 1} | \tilde{J}_Z | \psi_{\mu, 1} \rangle
=  - \frac{ F_\lambda }{16} - E_0 \cos 2 \theta_\mu,
\nonumber \\
& \langle \psi_{\mu, 2} | \tilde{J}_Z | \psi_{\mu, 2} \rangle
=  - \frac{ F_\lambda }{16} + E_0 \cos 2 \theta_\mu,
\nonumber \\
& \langle \psi_{\mu, 2} | \tilde{J}_Z | \psi_{\mu, 1} \rangle
= \langle \psi_{\mu, 1} | \tilde{J}_Z | \psi_{\mu, 2} \rangle
= E_0~( \pm \sin 2 \theta_\mu ).
\label{eqn:JZ}
\end{align}
\par

Finally, we provide a brief comment on a correlation effect arising from intersite
multipole--multipole interactions, including magnetic dipoles and octupoles other than
the quadrupoles discussed above.
This effect has not been considered in the mean-field treatment.
For the AFQ ordering, the effect of an AF correlation between the two sublattice sites manifests
as a modification of crystal field excitation, depending on the multipole--multipole coupling
strength.
Within the framework of our theory, an effective level shift of the excitation energy can be
incorporated into $\delta \varepsilon_\mu$ as a phenomenological parameter in
Eq.~(\ref{eqn:gamma2}).
Consequently, the AF correlation effect may also contribute to the reduction in transition
probability peak intensity discussed in Sect.~3.2.
\par

%%%%%%%%%%%%%%%%%%%%%%%%%%%%%%%%%%%%%%%%%%%%%%%%%%%%%%%%%%%%%%%%%%%%%%%%%%%%%%%%
\section{Case of Transverse BAW}
%%%%%%%%%%%%%%%%%%%%%%%%%%%%%%%%%%%%%%%%%%%%%%%%%%%%%%%%%%%%%%%%%%%%%%%%%%%%%%%%
By analogy with the longitudinal BAW discussed in Sect.~3.2, we consider a transverse BAW
vibrating in the $xy$-plane for $\bH \parallel [110]$.
The direction of displacement $u_{y'}$ is perpendicular to the BAW propagation direction ($x'$).
The three strain tensor components are replaced as
\begin{align}
\varepsilon_u  \rightarrow 0,~~
\varepsilon_v \rightarrow - 2 \varepsilon_{x' y'} \sin 2 \varphi,~~
\varepsilon_{xy} \rightarrow \varepsilon_{x' y'} \cos 2 \varphi.
\end{align}
For the transverse QS coupling, $| A_T |$ has the same form as Eq.~(\ref{eqn:ATf}), but it shows
a different $\varphi$ dependence:
\begin{align}
& a_L ( \varphi )
= \left[ 12 ( 4 - 3 \lambda ) g_c \cos 2 \varphi \right]^2  \langle \varepsilon_{x' y'}^2 \rangle, \\
& a_{YZ} ( \varphi ) = 48 ( g_b \sin 2 \varphi )^2 \langle \varepsilon_{x' y'}^2 \rangle.
\end{align}
As in Sect.~3.2, the same values are chosen for the parameters $\lambda$, $\bar{\omega}$, and
$g^2 \langle \varepsilon_{x' y'}^2 \rangle / \Delta^2$.
The PA-MAR transition probability $\bar{P}_{1 \rightarrow 2}^{(1,1)}$ shows sharp peaks when
$\varphi \rightarrow 0$, $\pi / 2$, and
$\pi$ in the normal phase ($\bar{\mu} = 0$), at which $a_{YZ} ( \varphi )$ approaches zero.
The induced ordered moment's increase in $\bar{\mu}$ causes an abrupt reduction in peak
intensity, while preserving the fourfold symmetry with respect to $\varphi$.
The fourfold symmetry is due to the vanishment of $\varepsilon_u$ associated with the QS
coupling ($g_a$) for the quadrupole component $O_u$, which is not generated by the transverse
BAW.
These characteristics of $\bar{P}_{1 \rightarrow 2}^{(1,1)}$ for the transverse BAW
propagating along the principal crystallographic axes differ from those for the longitudinal
BAW presented in Sect.~3.2.
This distinction arises from the $\varphi$ dependence of the transverse QS coupling ($A_T$).
\par

%%%%%%%%%%%%%%%%%%%%%%%%%%%%%%%%%%%%%%%%%%%%%%%%%%%%%%%%%%%%%%%%%%%%%%%%%%%%%%%%
\section{Displacement Vector of BAW Propagating in Cubic Lattice}
%%%%%%%%%%%%%%%%%%%%%%%%%%%%%%%%%%%%%%%%%%%%%%%%%%%%%%%%%%%%%%%%%%%%%%%%%%%%%%%%
In general, the displacement vector $\bu$ of a propagating BAW is expressed as a linear
combination of normal modes, each with its own distinct sound velocity.
When the propagation direction is rotated in the cubic $xy$-plane, the eigenvector
$\bu = (u_x, u_y)$ of a normal mode is neither parallel nor orthogonal to the wave number vector
$\bk = (k_x, k_y) = k ( \cos \varphi, \sin \varphi)$ for arbitrary values of $\varphi$ except for
$\varphi = 0$, $\pi / 4$, $\pi / 2$, $3 \pi / 4$, and so on.
This fact can be explained by solving the equations of motion for the BAW,
\cite{Kittel05}
\begin{align}
\left\{
\begin{array}{l}
\ds{ \rho \frac{ \partial^2 u_x }{ \partial t^2 } = C_{11} \frac{ \partial^2 u_x }{ \partial x^2 }
+ C_{44} \frac{ \partial^2 u_x }{ \partial y^2 }
+ ( C_{12} + C_{44} ) \frac{ \partial^2 u_y }{ \partial x \partial y } }, \\
\ds{ \rho \frac{ \partial^2 u_y }{ \partial t^2 } = C_{11} \frac{ \partial^2 u_y }{ \partial y^2 }
+ C_{44} \frac{ \partial^2 u_y }{ \partial x^2 }
+ ( C_{12} + C_{44} ) \frac{ \partial^2 u_x }{ \partial x \partial y } },
\end{array}
\right.
\label{eqn:motion}
\end{align}
using the mass density $\rho$ and the bulk elastic constants $C_{11}$, $C_{12}$, and $C_{44}$
for the cubic symmetry.
For a traveling BAW with frequency $\omega$, the components of the displacement vector are
given as $u_j = u_{j, 0} \exp [ i ( k_x x + k_y y - \omega t ) ]$ ($j = x, y$), where $u_{j, 0}$ is the
amplitude.
Substituting $u_j$ into Eq.~(\ref{eqn:motion}), we obtain
\begin{align}
\omega^2 \rho
\left(
\begin{array}{c}
u_x \\
u_y
\end{array}
\right) = \left(
\begin{array}{cc}
C_{11} k_x^2 + C_{44} k_y^2 & ( C_{12} + C_{44} ) k_x k_y \\
( C_{12} + C_{44} ) k_x k_y & C_{11} k_y^2 + C_{44} k_x^2
\end{array}
\right) \left(
\begin{array}{c}
u_x \\
u_y
\end{array}
\right),
\label{eqn:motion2}
\end{align}
which leads to the condition for the solution $(u_x, u_y) \ne (0, 0)$:
\begin{align}
& \left[ \omega^2 \rho - \frac{ k^2 }{2} ( C_+ + C_- \cos 2 \varphi ) \right]
\left[  \omega^2 \rho - \frac{ k^2 }{2} ( C_+ - C_- \cos 2 \varphi ) \right]
\nonumber \\
&~~~~~~
- \frac{ k^4 }{4} C_{2+}^2 \sin^2 2 \varphi = 0.
\end{align}
Here, the elastic constants are rewritten as $C_\pm \equiv C_{11} \pm C_{44}$ and
$C_{2+} \equiv C_{12} + C_{44}$.
Introducing an effective elastic coefficient
$C_\varphi \equiv ( C_-^2 \cos^2 2 \varphi + C_{2+}^2 \sin^2 2 \varphi )^{1/2}$, we obtain
the eigenvalues $\omega_\pm$ of frequency
\begin{align}
\omega_\pm = \sqrt{ \frac{ C_+ \pm C_\varphi }{2 \rho} } k.
\label{eqn:BAWom}
\end{align}
\par

We concentrate on the eigenvector $\bu$ corresponding to $\omega_+$ as a single normal mode.
Using the relationship between $C_-$, $C_{2+}$, and $C_\varphi$  with a new
parameter $\varphi_k$,
\begin{align}
C_- \cos 2 \varphi = C_\varphi \cos 2 \varphi_k,~~
C_{2+} \sin 2 \varphi = C_\varphi \sin 2 \varphi_k,
\label{eqn:Cphi}
\end{align}
we obtain $u_x \sin \varphi_k = u_y \cos \varphi_k$ from Eq.~(\ref{eqn:motion2}).
Since $\bk$ is parallel to the $x'$-axis, and $(u_{x'}, u_{y'})$ is related to $(u_x, u_y)$ by the
same transformation as given in Eq.~(\ref{eqn:rotation}), we obtain
\begin{align}
u_{x'} \sin ( \varphi_k - \varphi ) = u_{y'} \cos ( \varphi_k - \varphi ).
\label{eqn:uphi}
\end{align}
Combining Eqs.~(\ref{eqn:Cphi}) and (\ref{eqn:uphi}) and rewriting
$\delta \varphi = \varphi_k - \varphi$, we derive
\begin{align}
C_\varphi
\left(
\begin{array}{c}
\cos 2 \delta \varphi \\
\sin 2 \delta \varphi
\end{array}
\right) = \left(
\begin{array}{cc}
\cos 2 \varphi & \sin 2 \varphi \\
- \sin 2 \varphi & \cos 2 \varphi
\end{array}
\right)
\left(
\begin{array}{c}
C_- \cos 2 \varphi \\
C_{2+} \sin 2 \varphi
\end{array}
\right).
\end{align}
This leads to
\begin{align}
\tan 2 \delta \varphi = - \frac{ \Delta C \sin 4 \varphi }{ 1 + \Delta C \cos 4 \varphi },
\end{align}
where
\begin{align}
\Delta C \equiv \frac{ C_{11} - C_{12} - 2 C_{44} }{ C_{11} + C_{12} }.
\end{align}
An isotropic condition, $\Delta C = 0$, implies $\delta \varphi = 0$, indicating that
$\bu = ( u_{x'}, 0 )$ of a longitudinal BAW is parallel to the propagation direction ($x'$) for arbitrary
values of $\varphi$.
Conversely, for $\Delta C \ne 0$, this holds at $\varphi = 0$, $\pi / 4$, $\pi / 2$, $3 \pi / 4$, and so
on.
These values of $\varphi$ correspond to $[100]$, $[110]$, $[010]$, and $[\bar{1}10]$, respectively.
The analogous argument also applies to $\omega_-$ in Eq.~(\ref{eqn:BAWom}), and
$\bu = ( 0, u_{y'} )$ of a transverse BAW is orthogonal to the propagation direction ($x'$).
\par

%\vspace*{-0.2cm}
%%%%%%%%%%%%%%%%%%%%%%%%%%%%%%%%%%%%%%%%%%%%%%%%%%%%%%%%%%%%%%%%%%%%%%%%%%%%%%%%
%Macros for Supplemental Material
%%%%%%%%%%%%%%%%%%%%%%%%%%%%%%%%%%%%%%%%%%%%%%%%%%%%%%%%%%%%%%%%%%%%%%%%%%%%%%%%
%\renewcommand{\figurename}{Fig. S \hspace{-0.25cm}}
%\renewcommand{\theequation}{S\arabic{equation}}
%\renewcommand{\thesection}{S\arabic{section}}
%\setcounter{equation}{0}
%%%%%%%%%%%%%%%%%%%%%%%%%%%%%%%%%%%%%%%%%%%%%%%%%%%%%%%%%%%%%%%%%%%%%%%%%%%%%%%%

%%%%%%%%%%%%%%%%%%%%%%%%%%%%%%%%%%%%%%%%%%%%%%%%%%%%%%%%%%%%%%%%%%%%%%%%%%%%%%%%%%%%%%%%%%%%%%%%%%%

\end{document}